\documentclass[a4paper, 11pt]{article}
\usepackage[left=1.2in, right=1.2in, top=1.5in, bottom=1.5in]{geometry}
\usepackage[T1]{fontenc}
\usepackage[utf8]{inputenc}
\usepackage{amsmath, amssymb, amssymb, amsthm, mathrsfs}
\usepackage[colorlinks]{hyperref}
  \hypersetup{
    colorlinks,
    citecolor=magenta,
    linkcolor=blue,
    urlcolor=black}

\newtheorem{theorem}{Theorem}[section]
\theoremstyle{definition}
\newtheorem{definition}{Definition}[section]

\begin{document}

\begin{center}
  {\Large\bf Towards quantum field theory
  on spacetimes with boundaries}\\
  \vspace{12pt}
  
  {\footnotesize Gabriele Nosari}
  \vspace{5pt}
  
  {\footnotesize \it Dipartimento di Fisica, Universit\`a degli Studi di Pavia, Via Bassi, 6, I-27100 Pavia, Italy.\\
  Istituto Nazionale di Fisica Nucleare - Sezione di Pavia, Via Bassi, 6, I-27100 Pavia, Italy.\\
  Department of Mathematics, University of York, Heslington, York YO10 5DD, UK.\\
  gabriele.nosari@pv.infn.it}
\end{center}

\begin{abstract}
\noindent In this paper it is given an explicit construction of the algebraic quantization of a massless scalar field on two prototypical examples of spacetime with boundary, usually related to the Casimir effect. This paper is based on Ref.~\cite{cite:DNP14}.
\end{abstract}
\begin{center}
  {\footnotesize{\it Keywords:} Algebraic quantum field theory; Casimir effect; Spacetime with boundaries.}
\end{center}

\section{Introduction}

Algebraic quantization has been applied on the class of globally hyperbolic spacetime for many decades, leading to remarkable results. Nonetheless, the presence of a boundary calls for a separate treatment, since, in general, it breaks global hyperbolicity and it requires the introduction of boundary conditions to the field equation. Although a general theory for such situations is lacking, we develop an explicit construction of the algebraic quantization of a massless scalar field on two prototypical geometries, by exploiting the method of images. This well known technique gives a way to relate solutions of the boundary value problem to solution of the field equation on Minkowski spacetime, filling the gap with the standard theory. At the same time, we point out some limits of reliability of this scheme.

\section{Construction of algebra of observables}

A quantum field theory for a massless scalar field is built out of a Cauchy problem for the Klein-Gordon operator $P$, {\it i.e.}, for suitable initial data $f$, $g\in\mathrm C^\infty(\mathbb R^3)$ at a fixed time $t_0\in \mathbb R$,
  \begin{equation}\label{eqref:IV}
    \begin{cases}
      P\varphi=(-\partial^2_t+\nabla^2-\xi R)\varphi=0 \\
      \varphi|_{t_0}=f\,,\,\,\partial_t\varphi|_{t_0}=g,
    \end{cases}
  \end{equation}
where $\xi\ge0$ is a coupling with $R$, the Ricci scalar.\footnote{On any $(\mathcal U\subset\mathbb R^4, g)$, being $g$ the Minkowski metric (with signature $(3,1)$), $R\equiv0$; nonetheless, the coupling term plays a role in the definition of the stress-energy tensor $T_{\mu\nu}$.} Let $C^\infty_{0}(\mathbb R^4)$ and $C^\infty_{tc}(\mathbb R^4)$ be the set of smooth functions compactly supported and timelike compact, respectively.\footnote{We introduce a useful notation (which can be adapted to every set of function):
  \[
    \mathrm{C}^\infty_{tc}(\mathbb{R}^4; P) \equiv \frac{\mathrm{C}^\infty_{tc}(\mathbb{R}^4)}{P[\mathrm{C}^\infty_{tc}(\mathbb{R}^4)]}.
  \] }

On Minkowski spacetime there exist two operators, $E^\pm: C^\infty_{tc}(\mathbb R^4)\to C^\infty(\mathbb R^4)$, called {\it advanced} ($+$) and {\it retarded} ($-$) {\it fundamental solutions}, satisfying 
  \begin{enumerate}
    \item $P\circ E^\pm = E^\pm\circ P = \mathsf{id}|_{C^\infty_{tc}(\mathbb R^4)}$,
    \item $\forall \alpha\in C^\infty_{tc}(\mathbb R^4)$, $\mathrm{supp}(E^\pm(\alpha))\subseteq J^\mp(\mathrm{supp}(\alpha))$;
  \end{enumerate}
where $J^\pm(\mathcal O)$ indicate the causal future ($+$) and the causal past ($-$) of $\mathcal O\subset\mathbb R^4$. From $E^\pm$ we can construct the causal propagator $E\doteq E^+-E^-$, which induces an isomorphism between $\mathrm{C}^\infty_{tc}(\mathbb{R}^4; P)$ and $\mathcal S(\mathbb R^4)$, the space of smooth solutions to $P\varphi=0$.

We adopt Cartesian coordinates $(t,x,y,z)$, shortened as $(\underline x,z)$, and define:
  \[
    \mathbb H^4 \doteq \{x\in\mathbb R^4 | z\ge 0\};\quad\quad\quad
    Z \doteq \{x\in\mathbb R^4 | 0\le z\le d,\, d\in\mathbb R\}.
  \]
We use $\Omega$ as a label for either $\mathbb H^4$ or $Z$. In both cases, $\Omega$ is not a globally hyperbolic spacetime, meaning that an initial value problem like \eqref{eqref:IV} is ill-defined. This motivates us to consider a boundary value problem, for example a Dirichlet one:
  \begin{equation}\label{eqref:BV}
    \begin{cases}
      P\varphi=0 \quad\text{on}\,\,\Omega\\
      \varphi|_{\partial\Omega}=0.
    \end{cases}
  \end{equation}
We collect every smooth solution of the system in $\mathcal{S}^{D}(\Omega)$, which we call {\it space of dynamical field configurations}. Smoothness on $\Omega$ entails that we can wonder whether each element of $\mathcal{S}^D(\Omega)$ admits a smooth extension to smooth function to the whole Minkowski spacetime which is, moreover, a solution to the massless wave equation thereon. We define:
  \begin{eqnarray*}
    \mathrm{C}^\infty_{tc,-}(\mathbb{R}^4) &\doteq&
      \{\alpha\in\mathrm C^\infty_{tc}(\mathbb R^4)\,|\,\alpha(\underline x, z)=-\alpha(\underline x, -z)\};\\
    \mathrm{C}^\infty_{tc,Z}(\mathbb{R}^4) &\doteq&
      \{\alpha\in\mathrm C^\infty_{tc,-}(\mathbb R^4)\,|\,\alpha(\underline x, z)=-\alpha(\underline x,2d-z)\}.
  \end{eqnarray*}
Exploiting the properties of the causal propagator, we prove two useful facts:
  \[
    \mathcal{S}^{D}(\mathbb{H}^4)\simeq\mathrm{C}^\infty_{tc,-}(\mathbb{R}^4; P);\quad\quad\quad
    \mathcal{S}^{D}(Z)\simeq\mathrm{C}^\infty_{tc,Z}(\mathbb{R}^4; P).
  \]

Following the construction described in Ref.~\cite{cite:Book} (Ch.~3), and Ref.~\cite{cite:Ben14}, we identify as generators of the algebra of observables a separating and optimal set of linear functionals on $\mathcal{S}^{D}(\Omega)$.\footnote{For the definition of separating and optimal set see Ref.~\cite{cite:Ben14}} We call
  \begin{eqnarray}
    \eta&: \mathrm C^\infty_0(\mathbb R^4) \to \mathrm C^\infty(\mathbb R^4), \quad
      &\eta(f)(\underline{x},z)\doteq \frac{1}{\sqrt{2}}\biggl(\!f(\underline{x},z)-f(\underline{x},-z)\!\biggr); \label{eta}\\
    N &:\mathrm C^\infty_0(\mathbb R^4) \to \mathrm C^\infty(\mathbb R^4), \quad
      &N(f)(\underline{x},z)\doteq\!\!\!\sum\limits_{n=-\infty}^\infty\!\!\!\biggl(\!f(\underline{x},2nd+z)-f(\underline{x},2nd-z)\!\biggr)\label{N}
  \end{eqnarray}
the building blocks of the {\it method of images} for $\mathbb H^4$ and $Z$, and we define the two set of generators:
  \begin{itemize}
    \item $\mathcal{O}^{D}(\mathbb{H}^4)$ being the span of 
      $F_{[\zeta]}:\mathrm{C}^\infty_{tc,-}(\mathbb{R}^4; P) \to \mathbb{C}$, 
      with $[\zeta] \in \eta \left[ \mathrm{C}^\infty_{0}(\mathbb{R}^4; P) \right]$
      such that \[F_{[\zeta]}([\alpha])=\int_{\mathbb{H}^4}\,\zeta(x) E(\alpha)(x)\,d^4x\]
    \item $\mathcal{O}^{D}(Z)$ as above, {\it mutatis mutandis} (note in particular that $\eta$ is replaced by $N$).
  \end{itemize}
At this stage, the two cases, $\mathbb H^4$ and $Z$, diverge considerably. While $\eta \left[ \mathrm{C}^\infty_{0}(\mathbb{R}^4; P) \right]\subset\mathrm{C}^\infty_{0}(\mathbb{R}^4; P)$, the range of $N$ has a trivial intersection with compactly supported functions.

Given a set of generators, we can build a collection of multilinear functionals\footnote{The generated functionals are required to be {\it regular}, a condition on all functional derivatives $F^{(n)}$ which ensures \eqref{starprod} being well-defined. We omit all details here, but the interested reader should check in Ref.~\cite{cite:Book} at Ch.~2.} $\mathcal A^D(\Omega)$, defining observables for a massless scalar field on $\Omega$. We drop the $[h]$ subscript to denote a generic $F\in \mathcal A^D(\Omega)$. Assigning a $\ast$-involution on functionals as complex conjugation, the definition of a proper $\ast$-algebra of observables is a matter of endowing $\mathcal{A}^{D}(\Omega)$ with a suitable product. It is defined in terms of a power series, in which the zeroth order is the point-wise product, while the first order encodes the canonical commutation relations:
  \begin{equation}\label{starprod}
    F\star_\Omega F^\prime=\sum\limits_{n=0}^\infty\frac{i^n\hbar^n}{2^n n!}\langle F^{(n)},E^{\otimes n}_\Omega(F^{\prime (n)})\rangle, 
    \quad \forall F,\,F^\prime \in \mathcal A^D(\Omega),
  \end{equation} 
where $F^{(n)}$ is the $n^{\mathrm{th}}$ functional derivative of $F$, and where $E_\Omega$ is a suitable bidistribution. We define then:
  \begin{eqnarray*}
    E_{\mathbb{H}^4}&: \eta[\mathrm{C}^\infty_0(\mathbb{R}^4)]\to\mathcal{S}^{D}(\mathbb{H}^4), 
      \quad &E_{\mathbb{H}^4}\doteq\chi_{\mathbb{H}^4}(\eta\circ E);\\
    E_{Z}&: N[\mathrm{C}^\infty_0(\mathbb{R}^4)]\to \mathcal{S}^{D}(Z), 
      \quad &E_{Z}\doteq\chi_{Z}(N \circ E),
  \end{eqnarray*}
where $\chi_{\Omega}$ is the restriction map, while $E$ is the causal propagator of the wave operator on Minkowski spacetime. The two bidistributions replace the causal propagator in the definition of $\star$-product of the algebra of observables of the massless scalar field on Minkowski space -- see in Ref.~\cite{cite:Book} (Ch.~2). Remarkably, $\star_\Omega$ consists of a deformation of the (classical) point-wise product (technically, it is a {\it deformation quantization}). We can therefore define the {\it $\ast$-algebra of ``confined'' fields} on $\Omega$ $(\mathcal{A}^{D}(\Omega), \star_{\Omega}, \ast)$.
 
In order to investigate the structural property of the theory outlined, we can eventually state the following result:
\begin{theorem}\label{theorem1}
  $(\mathcal{A}^{D}(\Omega), \star_\Omega, \ast)$ is causal and it fulfils the time-slice axiom. Furthermore, it holds that $(\mathcal{A}^{D}(\Omega), \star_\Omega, \ast)$ and $(\mathcal{A}(\mathbb{R}^4), \star, \ast)$ are $\ast$-isomorphic, when restricted to any $\mathcal{U}\in\Omega$, open and globally hyperbolic.
\end{theorem}
\noindent Causality and the time slice axioms are two standard requirements of the algebraic quantization scheme -- see in Ref.~\cite{cite:Book} (Chs.~1 and 3). In this context, they could be understood in terms of the following sketch: The boundary or rather the Dirichlet boundary conditions are acting as a mirror. Every propagated initial data is ``bouncing'', without any absorption and, consequently, without any loss of information within its causal domain. At the same time Th.~\ref{theorem1} suggests that such behaviour is compatible with causality, as it shows that no observable can detect the presence of a boundary if it is localized away from it. This idea is inspired by that of {\it F-locality}, introduced first in Ref.~\cite{cite:Kay92}, and it goes in the direction of a recent proposal of a new axiomatization, the {\it principle of general local convariance} -- see in Ref.~\cite{cite:Book} (Ch.~4) and references therein.

\section{States and the method of images}

An algebraic state for $\mathcal A^{D}(\Omega)$ is any linear functional $\omega_\Omega:\mathcal A^D(\Omega)\to \mathbb C$ which is 
  \begin{enumerate}
    \item {\it positive}, $\omega_\Omega(F\star_\Omega F^\ast)\ge0$ $\forall F\in \mathcal A^D(\Omega)$,
    \item {\it normalized}, $\omega_\Omega(I)=1$, $I$ being the unit of the algebra.
  \end{enumerate}  
We need a counterpart on $\mathcal A^D(\Omega)$ of {\it Hadamard states}. On Minkowski spacetime, a Hadamard state is a quasi-free\footnote{A {\it quasi-free} state is a state which has odd $n$-point functions vanishing and every even $n$-point function is specified uniquely by its two-point function (further details in Ref.~\cite{cite:Book} at Ch.~5).} algebraic state whose two-point function $\omega_2\in \mathcal{D'}(\mathbb{R}^4\times\mathbb{R}^4)$ fulfils the microlocal spectrum condition ($\mu$SC)
  \begin{equation}\label{muSC}
    WF(\omega_2)=\left\{(x,x^\prime,k_x,-k_{x^\prime})\in T^*(\mathbb{R}^4\times\mathbb{R}^4)\setminus \{\mathbf 0\}\;|\;
    (x,k_x)\sim(x^\prime,k_{x^\prime}),\;k_x\triangleright 0\right\}  
  \end{equation}
where $x\sim x^\prime$ if they are connected by a null-like geodesic and $g^{-1}(k_{x^\prime})$ is the parallel transport of $g^{-1}(k_x)$ along it (here $g$ is the Minkowski metric). The singular structure is fixed only by the local properties of the geometry and of the differential problem. Locally it implies that a Hadamard state is defined by an integral kernel of the form 
  \begin{equation}
    \omega_2(x,x^\prime)=H(x,x^\prime)+W(x,x^\prime),
  \end{equation}
where $W(x,x^\prime)$ is a smooth function, while $H(x,x^\prime)$ is a singular term which does not depend on $\omega$, called {\it Hadamard parametrix} (see in Ref.~\cite{cite:Book} at Ch.~5). 

The following definition is inspired by Ref.~\cite{cite:Kay79}, by analogous choices in different contexts, e.g.~linearized gravity -- see in Ref.~\cite{cite:BDM14}, and by F-locality.
\begin{definition}\label{hadst}
  A state $\omega_\Omega:\mathcal{A}^{D}(\Omega)\to\mathbb{C}$ is said to be Hadamard if it is an Hadamard state in the usual sense 
  whenever restricted to any $\mathcal{U}\in\Omega$, globally hyperbolic.
\end{definition}
When $\Omega=\mathbb{H}^4$, we can give explicit examples of states. In fact, extending the domain of $\eta$ to $\mathrm C^\infty_{tc}(\mathbb R^4)$, it has a well-defined pull-back action on generators: 
  \[
    \eta^\ast F([f]) = F(\eta([f])), \quad\quad \forall \,F\in \mathcal{O}^{D}(\mathbb{H}^4), 
    \quad \forall \,[f]\in\mathrm{C}^\infty_{tc}(\mathbb{R}^4; P).
  \]
The domain of $\eta^\ast$ is extendible to $\mathcal A^{D}(\mathbb H^4)$ and it defines an injective \hbox{$\ast$-homomorphism} between $\mathcal{A}^{D}(\mathbb{H}^4)$ and $\mathcal{A}(\mathbb{R}^4)$. Thus, a state $\widetilde\omega_{\mathbb H^4}$ on $\mathcal{A}^{D}(\mathbb{H}^4)$ can be defined as
  \begin{equation}
    \widetilde\omega_{\mathbb H^4}(F)\doteq\omega(\eta^\ast(F)), \quad\quad \forall F\in \mathcal A^D(\mathbb H^4),
  \end{equation}
for every Hadamard state $\omega$ on $\mathcal{A}(\mathbb{R}^4)$\footnote{The procedure is injective but not surjective.}. Furthermore it is Hadamard according to Def.~\ref{hadst}. We call it {\it image state}, since the integral kernel of the two-point function can be built by means of the method of images.

For $\Omega=Z$ limitations on the feasibility of this method occur. The action of $N$, defined in \eqref{N}, on a given two-point function $\omega_2(x, x^\prime)$ is defined as follows
  \begin{equation*}
    \left(N\otimes \mathsf{id}\right)\,\omega_2(x,x^\prime)\doteq\!\!\!
    \sum\limits_{n=-\infty}^\infty\!\! \biggl(\omega_2(\underline{x},2nd+z, x^\prime)-\omega_2(\underline{x},2nd-z, x^\prime)\!\biggr),
  \end{equation*}
where the series might not converge. A sufficient condition for convergence is given in Ref.~\cite{cite:DNP14}. It holds, moreover, that:
\begin{theorem}\label{imagetroubles}
  Given a Hadamard state $\omega$ on $\mathcal A(\mathbb R^4)$ whose image series converges, by the method of images it induces a Hadamard state $\widetilde\omega_{\mathring Z}$ defined on $\mathcal A^{D}(\mathring Z)\subset\mathcal A^{D}(Z)$, the $\ast$-subalgebra generated by $\mathcal O^{D}(\mathring Z)$, the set of linear functionals labelled by $\eta[\mathrm{C}^\infty_0(\mathring Z; P)]$. 
\end{theorem}
\noindent The method of images is applied to derive the well known non-vanishing vacuum energy density, responsible for the Casimir effect (see in Ref.~\cite{cite:Mil01} and references therein). The limitation to observables in the interior of $Z$ might suggest that such computations do not tell the whole story, at least when reaching the boundary. This point should be object of further inspections.

\section{Extended algebra of regularized observables}\label{extension}

The great improvement of the functional approach consists of providing a handy method to build the extended algebra of regularized observables (for references see in Ref.~\cite{cite:Book} at Ch.~2). Within such a framework, Wick polynomials are replaced by the notion of {\it microcausal ($\mu$-) functionals}, which extends the class of observables generated by linear functionals. We define a $\mu$-functional as a smooth functional $F:\mathrm C^\infty(\mathbb R^4)\to\mathbb R^4$ such that for all $n\in \mathbb N$ the $n^\mathrm{th}$ functional derivative $F^{(n)}$ fulfils
  \[
    WF(F^{(n)}) \cap \left(\overline{V}^n_+\cup\overline{V}^n_-\right) = \emptyset
  \] 
where $\overline{V}_\pm$ are the subsets of $T^*\mathbb R^4$ formed by elements $(x_i,k_i)$ where each covector $k_i$, $i=1,...,n$ lies in the closed future ($+$) and in the closed past ($-$) light cone. Barring technicalities (see in Ref.~\cite{cite:DNP14}), this definition transfers to $\Omega$. Furthermore, \mbox{$\mu$-functionals} have finite expectations values under the action of Hadamard states of Def.~\ref{hadst}.

On the new class of functionals the product $\star$ is not well-defined. Thus we replace $E$ with $-2iH$ in \eqref{starprod}, where $H$ is the Hadamard parametrix, obtaining a new product $\star_H$, which is well-defined on $\mu$-functionals and preserves the canonical quantization. We call $(\mathcal A_\mu(\mathbb R^4), \star_H, \ast)$ the {\it extended algebra of observables}. On $\Omega$ this deformation does not work globally. From the analysis of {\it image states} we get an important insight on the global structure of the two-point function:
  \begin{eqnarray*}
    WF(\widetilde\omega_{\mathbb H^4})&=\left\{(x,x^\prime,k_x,-k_{x^\prime})\in T^*(\mathbb{H}^4\times\mathbb{H}^4)\setminus \{\mathbf 0\}
    \;|\;(x,k_x)\sim_-(x^\prime,k_{x^\prime}),\;k_x\triangleright 0\right\} \\
    WF(\widetilde\omega_{Z})&=\left\{(x,x^\prime,k_x,-k_{x^\prime})\in T^*(Z\times Z)\setminus \{\mathbf 0\}
    \;|\;(x,k_x)\sim_Z(x^\prime,k_{x^\prime}),\;k_x\triangleright 0\right\} 
  \end{eqnarray*} 
where $x\sim_-x^\prime$($x\sim_Zx^\prime$) if $x$ or its reflection(s) on $\partial\mathbb H^4$($\partial Z$) are related to $x^\prime$ or its reflection(s) on $\partial\mathbb H^4$($\partial Z$) by $\sim$, defined in \eqref{muSC}. Observe that the algebra, built out of the deformed product $\star_H$ is well-defined only locally, in each globally hyperbolic subregion of $\Omega$. Globally we face an obstruction related to the additional singularities due to the reflections at the boundary. In Ref.~\cite{cite:DNP14} it has been shown that this problem can be circumvented by means of a further deformation of the extended algebra of observables, yielding an extended algebra of regularized observables $(\mathcal A^D_\mu(\Omega), \star_\Omega, \ast)$. In this framework, the local energy density does not coincide with the Wick ordered one, which has a vanishing expectation value on the vacuum state. Furthermore, the definition of $\star_\Omega$ is {\it non local}, since it encodes an information of a topological feature, {\it i.e.}, $\partial\Omega\neq0$.

\section*{Acknowledgements}

This work has been supported by a Ph.D.~grant of the University of Pavia. The author is grateful to the Department of Mathematics of the University of York for the kind hospitality during the preparation of this manuscript. A special thanks to C.~Fewster for his comments and for suggesting the deformation approach in \ref{extension}, and to both C.~Dappiaggi and N.~Pinamonti for their comments and the careful review of this paper.

\end{document}